\documentclass[preprint,aps,superscriptaddress,nofootinbib]{revtex4-1}
\usepackage{graphicx}
\usepackage[toc,page]{appendix}
\usepackage{braket}
\newcommand{\del}{\partial}

\begin{document}
\title{New solutions in pure gravity with degenerate tetrads
}
\author{Romesh K. Kaul}
\email{kaul@imsc.res.in}
\affiliation{The Institute of Mathematical Sciences, Chennai-600113, INDIA}

\author{Sandipan Sengupta}
\email{sandipan@phy.iitkgp.ernet.in}
\affiliation{Department of Physics and Centre for Theoretical Studies, Indian Institute of Technology Kharagpur, Kharagpur-721302, INDIA}

\begin{abstract}
In first order formulation of pure gravity, we find a new class of solutions 
to the equations of motion  represented by degenerate four-geometries.   These 
configurations are described by non-invertible tetrads with two zero eigenvalues 
and admit non-vanishing torsion. The homogeneous ones among these infinitely many degenerate 
solutions admit a geometric classification provided 
by the three fundamental geometries that a closed two-surface can accomodate, namely, $E^2$, $S^2$ and $H^2$.


\end{abstract}
  
\maketitle

\section{Introduction}
Classical theory of gravity admits two different descriptions. The usual one is 
based on the metric (second order) formulation, where the action functional is 
written in terms of the metric fields $g_{\mu\nu}(x)$:
\begin{eqnarray}\label{EH}
S[g]~=~\frac{1}{2\kappa^2}~\int d^4 x~ \sqrt{g} g^{\mu\nu} R_{\mu\nu}(g)
\end{eqnarray}
The equations of motion resulting from the variation of this action leads to 
the theory of Einstein gravity. 
There exists another formulation, known as the first order theory, where 
Lorentzian (Euclidean) gravity can manifestly be given a local $SO(3,1)$ 
($SO(4))$ gauge theoretic interpretation. The corresponding action functional 
depends on two sets of fields, namely, tetrad $e_\mu^I(x)$ and connection 
$\omega_{\mu}^{~IJ}(x)$:
\begin{eqnarray}\label{HP}
S[e,\omega]~=~\frac{1}{8\kappa^2}~\int d^4
 x~\epsilon^{\mu\nu\alpha\beta}\epsilon_{IJKL}
e_{\mu}^I e_{\nu}^J R_{\alpha \beta}^{~~ KL}(\omega)
\end{eqnarray}
where $R_{\mu \nu}^{~IJ} (\omega)=\del_{[\mu}
\omega^{~IJ}_{\nu]}+\omega^{~IK}_{[\mu}\omega^{~KJ}_{\nu]}$ 
is the field strength of the gauge connection
$\omega_{\mu}^{~IJ}$ of the gauge group.
These two descriptions of gravity theory are equivalent only when the tetrad 
(or metric) is invertible, implying $\det e_\mu^I \neq 0$. This is so because 
the second order action (\ref{EH}) requires the inverse metric $g^{\mu\nu}$ in 
its construction, whereas the first order action (\ref{HP}) does not. In other 
words, the first order theory can admit solutions of equations of motion 
corresponding to degenerate tetrads, which are not perceived at all by the 
metric formulation.

There has been quite a lot of discussion on degenerate spacetime geometries in
the literature, both in the context of metric \cite{einstein,hawking} and 
tetrad  \cite{regge,tseytlin,jacobson,madhavan,bengtsson,kaul} formulations. 
In first order gravity, explicit examples of degenerate tetrad 
solutions of the equations of 
motion were first constructed by Tseytlin \cite{tseytlin}. In a more recent 
work \cite{kaul}, a general framework to obtain all possible solutions with 
degenerate tetrads having one zero eigenvalue was set up. There, it was also 
shown that this space of solutions contains a special class that corresponds to 
the eight fundamental homogeneous three-geometries  as classified by Thurston 
\cite{thurston}.

However, there exists another nontrivial case that deserves special attention, 
namely, first order gravity theory for degenerate tetrads with two zero 
eigenvalues. This is  what we focus on here. As elucidated in the next 
few sections, the solution space for this theory possesses a rich structure and 
the essential details are qualitatively different from the case with one zero 
eigenvalue in quite a few respects. The only other case of possible interest, corresponding to tetrads with three null eigenvalues, is also discussed briefly.

As emphasized in \cite{kaul}, it is important to realize that 
 these degenerate tetrad solutions in pure 
gravity correspond to an unusual causal structure of spacetime 
\cite{sengupta}. These  could also mediate topology change \cite{horowitz}.  
Further, each of these solutions provides  a saddle point of the quantum 
path integral in first order gravity. Hence, it is important to obtain 
all  such solutions and analyze their properties to unravel their 
potential role in classical as well as quantum gravity. 
\section{Degenerate tetrad with two zero eigenvalues}
Variations of the first order action functional for 
four-dimensional Euclidean  gravity  (\ref{HP})  with respect to the 
fields $\omega_{\beta}^{~KL}$ and $e_\beta^L$ result in the 
following equations of motion   respectively: 
\begin{eqnarray}\label{eom1}
e_{[\mu}^{[I} D^{}_{\nu}(\omega)e_{\alpha]}^{J]}~&=&~0\\
e_{[\mu}^{[I} R_{\nu\alpha]}^{~~JK]}(\omega)~&=&~0\label{eom2}
\end{eqnarray}
Let us now consider degenerate tetrad fields of the form: 
\begin{eqnarray}\label{tetrad}
e_\mu^I~=~\left(\begin{array}{cccc}
e_x^1 & e_y^1 & 0 & 0\\
e_x^2 & e_y^2 & 0 & 0\\
0 & 0 & 0 & 0\\
0 & 0 & 0 & 0
\end{array}\right) \label{degenerate}
\end{eqnarray}
 with $\mu\equiv(x,y,z,\tau)$ and $I\equiv(1,2,3,4)$.
The two null eigenvalues of the tetrad have 
been chosen to lie along the $z$ and $\tau$ directions, whereas the remaining 
coordinates $x,y$ span a (reduced) two dimensional subspace with a 
non-degenerate two-metric. Non-zero determinant  of the corresponding diad 
is $e\equiv e^1_x e^2_y - e^2_x e^1_y= \frac{1}{2}\epsilon^{ab}\epsilon_{ij}
e_a^i e_b^j$ $(a\equiv x,y;~i\equiv 1,2)$.
We shall denote the inverse of this diad by $e^a_i$: $ ~e^a_i e^j_a =
 \delta^j_i$ and $ e^a_i e^i_b = \delta ^a_b$. 
It is always possible to reduce any arbitrary tetrad with two zero eigenvalues 
to the form (\ref{degenerate}) above using local orthogonal rotations and general coordinate 
transformations. 
 
 Twenty four connection equations of motion in (\ref{eom1}) can be 
 separated into (6+6+12) equations as: 
\begin{eqnarray}\label{eom3}
e_{[z}^{[I} D^{}_{a}(\omega)e_{b]}^{J]}~&=&~0\\\label{eom4}
e_{[\tau}^{[I} D^{}_{a}(\omega)e_{b]}^{J]}~&=&~0\\\label{eom5}
e_{[z}^{[I} D^{}_{\tau}(\omega)e_{a]}^{J]}~&=&~0
\end{eqnarray}
A similar decomposition of  16  tetrad equations of motion in (\ref{eom2}) leads to ($4+4+8$) equations:
\begin{eqnarray}\label{eom6}
e_{[z}^{[I} R_{ab]}^{~~JK]}(\omega)~&=&~0\\\label{eom7}
e_{[\tau}^{[I} R_{ab]}^{~~JK]}(\omega)~&=&~0\\\label{eom8}
e_{[z}^{[I} R_{\tau a]}^{~~JK]}(\omega)~&=&~0
\end{eqnarray}
In the following section, we find the most general solutions to these equations 
of motion.

\section{General solutions}
Let us begin our discussion with equation (\ref{eom3}), the first among the set 
of connection equations of motion. The $I=i,~J=3$ components of these equations 
lead to: ~$e^{[i}_a e^{j]}_b \omega_z^{~3j}=0,~$ which implies: 
$\omega_z^{~3j}=0$. Similarly, the $I=i,~J=4$ components of Eq.(\ref{eom3}) 
imply that $\omega_z^{~4j}=0$. Next, for $I=i,J=j$, we obtain:
\begin{eqnarray*}
e_{a}^{[i} D_z e_{b}^{j]}-e_{b}^{[i} D_z e_{a}^{j]}&=&0~,\\
or, ~~ e^a_i \partial_z e_a^i&=&0~\\
or, ~~  \del_z e&=&0
\end{eqnarray*}
which implies that the 2-determinant $e$ is $z$-independent. The only remaining 
equation from (\ref{eom3}), given by $I=3,~J=4$, is satisfied identically and 
hence does not lead to any new constraint. Altogether, Eqn.(\ref{eom3}) implies five constraints:
\begin{eqnarray}\label{eom9}
\omega_z^{~3j}=0,~~~~~\omega_z^{~4j}=0,~~~~~\del_z e=0~. 
\end{eqnarray}
Proceeding similarly with six equations in (\ref{eom4}) where again 
one equation is trivial for tetrads (\ref{degenerate}), we obtain five constraints from 
the other nontrivial equations:
\begin{eqnarray}\label{eom10}
\omega_\tau^{~3j}=0,~~~~~\omega_\tau^{~4j}=0,~~~~~\del_\tau e=0~.
\end{eqnarray}
Thus from 12 equations in (\ref{eom3}) and ({\ref{eom4}), we obtain ten 
constraints listed in Eqns.(\ref{eom9}) and (\ref{eom10}).
The last of the connection equations of motion (\ref{eom5}) is 
satisfied identically for the tetrad fields (\ref{tetrad}).
This completes our discussion of the connection equations of motion
(\ref{eom3}-\ref{eom5}).
Eight connection components, $ \omega_z^{~3i}, $ $  
\omega_z^{4i}$, $  \omega_\tau^{~3i}$ and 
$\omega_\tau^{~4i}$, are fixed  as  in Eqns.(\ref{eom9}) and 
(\ref{eom10}). Of the rest 16 connection components, as we shall see below,
some  will get fixed by the tetrad equations of motion 
 (\ref{eom6}-\ref{eom8}) which we study next.
 
 It is straight forward to check that the four tetrad equations  of motion 
 in (\ref{eom6}), for the choices $(I=i,~ J=3,~K=4), ~(I=i, ~J=j,~ K=3)$ and $(I=i, ~
J=j,~ K=4)$,   are respectively equivalent to the following set of 
$(2 + 1+ 1)$ equations:
\begin{eqnarray}\label{eom11}
R_{za}^{~~34}=0,~~~~~~e^a_i R_{za}^{~~3i}=0,~~~~~~e^a_i R_{za}^{~~4i}=0~.
\end{eqnarray}
The first subset  of 
two equations here  leads to:
\begin{eqnarray*}
\del_{[a}\omega_{z]}^{~34}=0
\end{eqnarray*}
These can be solved for the connection components  $\omega_z^{~34}$ and 
$\omega_{a}^{~34}$, implying that these are just pure gauge:
\begin{eqnarray}\label{w0}
\omega_z^{~34}=\del_{z}\Omega(x,y,z,\tau),~~~~~~\omega_a^{~34}=\del_{a}
\Omega(x,y,z,\tau)~.
\end{eqnarray}
In order to solve the last two equations of motion in (\ref{eom11}), let 
us introduce a parametrization of the four components of 
$\omega_a^{~j3}~(\omega_{a}^{~j4})$ in terms of a symmetric 
matrix $M^{kl}~(\bar{M}^{kl})$ with three independent components 
and a scalar $\phi~(\bar{\phi})$:
\begin{eqnarray}\label{Omega0}
\omega_a^{~j3}=e_a^l M^{jl}+\epsilon^{jl} e_a^l \phi,~~~~~~
\omega_a^{~j4}=e_a^l \bar{M}^{jl}+\epsilon^{jl} e_a^l \bar{\phi}
\end{eqnarray}
 Since both sets of fields $(M^{kl}, ~\phi)$ and $~(\bar{M}^{kl},
  ~{\bar \phi})$ and are arbitrary, this
 parametrization   does not imply any loss of generality. Using these in 
 the last two equations of   (\ref{eom11}), the most general solution for 
 the two connection components $\omega_{z}^{~34}$ and $\omega_{z}^{12}$ are 
 found to be:
  \begin{eqnarray}\label{w}
\omega_z^{~34}=\frac{e^a_j \left[\phi \del_z \bar{M}^{j}_a-\bar{\phi} \del_z 
M_a^j\right]}{\phi M^{kk}+\bar{\phi}\bar{M}^{kk}}~,~~~
\epsilon_{ij}\omega_z^{~ij}=\frac{e^a_j \left[M^{kk} \del_z 
M^{j}_a+\bar{M}^{kk} \del_z \bar{M}_a^j\right]}{\phi 
M^{kk}+\bar{\phi}\bar{M}^{kk}}~+~\epsilon_{ij}e^a_i \del_{z} e_a^j~
\end{eqnarray}
 where $M^j_a \equiv M^{jl} e^l_a$ and ${\bar M}^j_a \equiv {\bar M}^{jl} e^l_a$.  
 Comparing Eq.(\ref{w0}) with (\ref{w}), we note that the gauge function 
 $\Omega(x,y,z,\tau)$ is not independent, but rather is related to $M^{kl}, 
 ~\bar{M}^{kl},~\phi$ and $\bar{\phi}$ as:
\begin{eqnarray}\label{Omega}
\Omega(x,y,z,\tau)=\int^z dz ~\frac{e^a_j \left[\phi \del_z 
\bar{M}^{j}_a-\bar{\phi} \del_z M_a^j\right]}{\phi 
M^{kk}+\bar{\phi}\bar{M}^{kk}}~~.
\end{eqnarray}
 
We now turn to solve the four equations in (\ref{eom7}), which can be 
recast in terms of  an equivalent set of  ($2+1+1$) equations as:
\begin{eqnarray}\label{eom12}
R_{\tau a}^{~~34}=0,~~~~~~~e^a_i R_{\tau a}^{~~3i}=0,~~~~~~~
e^a_i R_{\tau a}^{~~4i}=0~.
\end{eqnarray}
 The first subset of two equations here implies that 
 $\omega_{\tau}^{~34}$ is  pure gauge, given by 
the same gauge function $\Omega(x,y,z,\tau)$ introduced in Eq.(\ref{w0}):
\begin{eqnarray}\label{eom13}
\omega_{\tau}^{~34}=\del_\tau \Omega(x,y,z,\tau)~.
\end{eqnarray}
 Using the parametrization introduced in (\ref{Omega0}), the remaining last 
 two  equations in (\ref{eom12}) can be solved for the two connection 
 components $\omega_{\tau}^{~34}$ and $\omega_{\tau}^{~12}$ as:
 \begin{eqnarray}\label{w2}
\omega_\tau^{~34}=\frac{e^a_j \left[\phi \del_\tau \bar{M}^{j}_a-\bar{\phi} 
\del_\tau M_a^j\right]}{\phi M^{kk}+\bar{\phi}\bar{M}^{kk}}~,~~~
\epsilon_{ij}\omega_\tau^{~ij}=\frac{e^a_j \left[M^{kk} \del_\tau 
M^{j}_a+\bar{M}^{kk} \del_\tau \bar{M}_a^j\right]}{\phi 
M^{kk}+\bar{\phi}\bar{M}^{kk}}~+~\epsilon_{ij}e^a_i \del_{\tau} e_a^j~
\end{eqnarray} 
 Finally, Eqns.(\ref{eom13}) and (\ref{w2})  along with Eqn.(\ref{Omega}) 
 imply: 
 \begin{eqnarray}\label{w3}
\Omega(x,y,z,\tau)~=~\int^\tau d\tau ~\frac{e^a_j \left[\phi \del_\tau 
\bar{M}^{j}_a-\bar{\phi} \del_\tau M_a^j\right]}{\phi 
M^{kk}+\bar{\phi}\bar{M}^{kk}}~=~\int^z dz ~\frac{e^a_j \left[\phi \del_z 
\bar{M}^{j}_a-\bar{\phi} \del_z M_a^j\right]}{\phi 
M^{kk}+\bar{\phi}\bar{M}^{kk}}
\end{eqnarray}
 With these, the remaining set of eight equations of motion contained in 
 (\ref{eom8}) reduce to identities, without leading to any further constraints. 
 This completes our analysis of the tetrad equations of motion 
 (\ref{eom6}-\ref{eom8}). 
  
 Of twenty four components of the connection $\omega_\mu^{~IJ}$, connections
  equations of motion (\ref{eom3}-\ref{eom5}) fix eight, 
$ \omega_z^{~3i}, $ $  \omega_z^{4i}$, $  \omega_\tau^{~3i}$ and 
$\omega_\tau^{~4i}$, to be zero as in Eqns.(\ref{eom9}) and 
(\ref{eom10}). Of rest 16 components,  tetrad equations of motion 
 (\ref{eom6}-\ref{eom8}) fix fourteen, $\omega_z^{~34}$,  $\omega_\tau^{~34}$,  
 $\omega_a^{~34}$, $\omega_a^{~3i}$, $ \omega_a^{~4i}$, $\omega_z^{~ij}$
 and $\omega_\tau^{~ij}$, which are completely
 characterized in terms of eight fields represented by two symmetric
 $M^{ij}$ and ${\bar M^{ij}}$ and two scalars $\phi$ and $\bar \phi$ through 
 Eqns.(\ref{w0}-\ref{w}) and  (\ref{eom13}-\ref{w3}).
 Although we have analyzed all the equations of motion, two connection 
components, $\omega_a^{~ij}$ $(a\equiv x,y;~i\equiv 1,2)$, are still left 
undetermined. This follows from the fact that the possible terms  containing the 
connection components $\omega_a^{~ij}$ in the equations of motion 
(\ref{eom3}-\ref{eom8}) are zero for the tetrads as represented in (\ref{degenerate}). This represents  an important difference from the case
where the tetrads have one zero eigenvalue \cite{kaul}. There, for tetrads 
with one null direction, three of the nine components of $\omega_a^{~ij}$ 
$(a\equiv x,y,z;~i\equiv 1,2,3)$ are determined (zero) where as six are left 
undetermined.
Now, in the present case with tetrads of two zero eigenvalues, 
the two arbitrary components of $\omega_a^{~ij}$ may be parametrized in terms 
of an arbitrary  vector field $N_a\equiv[N(x,y,z,\tau),~\bar{N}
(x,y,z,\tau)]$ as:
\begin{eqnarray}\label{K}
\omega_a^{~ij}=\bar{\omega}_a^{~ij}(e)+\epsilon^{ij}N_a~~~~~[a\equiv 
x,y;~i\equiv 1,2]
\end{eqnarray}
 where $\bar{\omega}_a^{~ij}(e)=\frac{1}{2}\left[e^b_i\del^{}_{[a}e_{b]}^j
-e^b_j\del^{}_{[a}e_{b]}^i -  e_a^l e^b_i e^c_j
\del^{}_{[b}e_{c]}^l\right]$ is the torsion-free spin-connection completely 
determined by the diads $e_a^i$.
The  ten independent fields $N$ and $\bar N$,  $(M^{kl}, ~\phi)$ 
and $(\bar{M}^{kl}, ~\bar{\phi})$, represent contorsion.  Thus, in
general, torsion is non vanishing  for the  solutions discussed here.

Since $R_{z\tau}^{~~34}(\omega)$ is zero for all
the above degenerate spacetimes, the action (\ref{HP}) is also zero 
for these solutions of equations of motion.
 
 The analysis presented in this article can be easily extended to study the solution space for tetrads with three zero eigenvalues, which is the only remaining case of possible interest. Such tetrad fields $e^I_\mu$ can be organized  by 
local orthogonal rotations and general coordinate transformation 
to have just one non-zero component, say $e^1_x$. For 
these, the connection equations of motion (\ref{eom3}-\ref{eom5}) are identically satisfied. 
Thus these yield no constraints.
On other hand equations of motion (\ref{eom6}-\ref{eom8}) yield the constraints:
$ R_{yz}^{~~23} = R_{yz}^{~~24} =R_{yz}^{~~34} =R_{y\tau}^{~~23} =
R_{y\tau}^{~~24} =R_{y\tau}^{~~34} =R_{z\tau}^{~~23} =R_{z\tau}^{~~24} =
R_{z\tau}^{~~34} =0$. While these can be solved for  
the connection components involved in these field strength 
components, other connection components are left undetermined.
Notably, there are no analogue of 
the contortion fields $N_a$ in this case and hence the solution space 
hardly exhibits any nontrivial structure.
 
 This completes our discussion  of the most general solutions of first order 
 gravity theory in four dimensions with degenerate tetrads having two or more zero 
 eigenvalues.  For the case with two null eigenvalues, there are an infinite number of such configurations in general, 
parametrized by the diads $e_a^i$ and the set of six arbitrary fields, 
$M^{kl},~\bar{M}^{kl},~\phi,~\bar{\phi},~N ~\mathrm{and} ~\bar{N}$. 
An interesting subset of these
degenerate space-time solutions consists of those  
 based on the types of fundamental homogeneous two-geometries 
represented by the diads $e_a^i$. It is well-known that there are only three 
independent homogeneous geometric structures that any closed two-surface can 
admit. These are  given by the Euclidean plane $E^2$, the two sphere $S^2$ 
and the hyperbolic plane $H^2$ \cite{thurston}. In the next 
section, we construct explicit four-dimensional torsional solutions 
(of first-order gravity theory) corresponding to these three two-geometries. 
Solutions based on more nontrivial two-geometries (e.g. non-orientable 
ones) can also be constructed by making 
appropriate choices of boundary conditions on the basic fields $e_\mu^I$ and 
$\omega_\mu^{~IJ}$. We do not discuss such examples here.

\section{Explicit solutions corresponding to three model two-geometries} 
 
 \subsection*{(i) $E^2$ geometry:}
Let us consider a case where the non degenerate two-subspace described by the 
diads $e_a^i$ in (\ref{tetrad}) is the Euclidean plane. The corresponding 
metric reads    
$ds_{(4)}^2=dx^2+dy^2+0+0$. The torsion-free spin-connection 
$\bar{\omega}_a^{~ij}(e)$ vanishes and hence the associated two-curvature
${\bar R}_{ab}^{~~ij} (\bar \omega)\equiv$ $\partial^{}_{[a}
 {\bar\omega}_{b]}^{ij}  + {\bar \omega}_{[a}^{~ik} {\bar 
 \omega}_{b]}^{~kj} =0$ and scalar two-curvature $\bar{R}
 (\bar{\omega}):=e^a_i e^b_j \bar{R}_{ab}^{~ij}(\bar{\omega})=0$. 
 
In general,  the six  fields $M^{kl} $ and $\bar{M}^{kl} $ 
in Eqn.(\ref{Omega0}) are arbitrary functions of the coordinates 
$(x,~y,~z,~\tau)$.  As an example, let us make  a simple choice  for these:
\begin{eqnarray}
M_a^k=\lambda e_a^{k},~ ~~~~~\bar{M}_a^{k}=\bar{\lambda} e_a^{k}
\end{eqnarray}
where $\lambda,\bar{\lambda}$ are arbitrary constants. This choice, when 
inserted into the general solutions above, leads to:
\begin{eqnarray*}
\Omega(x,y,z,\tau)=0,~~~~~\omega_z^{~12}=0=\omega_\tau^{~12},~~~~
~\omega_z^{~34}=0=\omega_\tau^{~34},~~~~~\omega_a^{~34}=0~.
\end{eqnarray*}
Finally, using the parametrization (\ref{Omega0}) and (\ref{K}) for 
the contorsion fields, all the connection one-forms are given by:
\begin{eqnarray*}
\omega^{13}&=&\lambda dx+\phi dy,~~~~~~\omega^{23}=-\phi dx+\lambda dy,~\\
\omega^{14}&=&\bar{\lambda} dx+\bar{\phi} dy,~~~~~~\omega^{24}=-\bar{\phi} 
dx+\bar{\lambda} dy,\\
\omega^{12}&=&N dx+\bar{N} dy, ~~~~~\omega^{34} =0
\end{eqnarray*}
where $\phi,~\bar{\phi},~N$ and $\bar{N}$ are arbitrary  functions of the 
spacetime coordinates ($x,y,z,\tau$).

\subsection*{(ii) $S^2$ geometry:}
Our next example is based on a degenerate four-metric, which contains a non 
degenerate $S^2$ subspace:
\begin{eqnarray*}ds_{(4)}^2=\ell^2\left[d\theta^2+
sin^2\theta d\chi^2\right]~+~0~+~0
\end{eqnarray*}
where $\theta$ and $\chi$ are  two angular coordinates on the two-sphere.
The only non-vanishing component of the torsion-free spin-connection 
is given by 
$\bar{\omega}_\chi^{~12}(e)=-cos\theta $ and the associated field strength 
also has only one non-vanishing component:
$\bar{R}_{\theta\chi}^{~~12}=\sin\theta$.  The two-metric characterizing the 
$S^2$ space has constant positive scalar curvature: $\bar{R}
(\bar{\omega})\equiv e^a_i e^b_j {\bar R}_{ab}^{~~ij}(\bar{\omega})
=\frac{2}{\ell^2}$. For the simple choice $M_a^k=\lambda 
e_a^{k},$ $~ \bar{M}_a^{k}=\bar{\lambda} e_a^{k}$ with $\lambda$ 
and $\bar{\lambda}$ as constants, the  solution of four dimensional 
gravity theory is given by the following   connection one-forms:
\begin{eqnarray*}
\omega^{13}&=&\ell\left[\lambda d\theta+\phi \sin\theta 
d\chi\right],~~~~~\omega^{23}=\ell\left[-\phi d\theta+\lambda \sin\theta 
d\chi\right]~,\\
\omega^{14}&=&\ell\left[\bar{\lambda} d\theta+\bar{\phi} \sin\theta 
d\chi\right],~~~~\omega^{24}=\ell\left[-\bar{\phi} d\theta+\bar{\lambda} \sin\theta 
d\chi\right]~,\\
\omega^{12}&=&N d\theta+[\bar{N}-\cos\theta] d\chi,~~~\omega^{34}=0~~.
\end{eqnarray*} 

\subsection*{(iii) $H^2$ geometry:}
The last of the three fundamental two-geometries is represented by the 
hyperbolic plane, nested within the four dimensional spacetime through a 
degenerate metric of the form:
\begin{eqnarray*}
ds_{(4)}^2=\frac{\ell^2}{y^2}(dx^2+dy^2)+0+0,~~~~~~~y>0 
\end{eqnarray*}
The torsion-free connection has only one non-vanishing component:
$\bar{\omega}^{~12}_x(e)=-\frac{1}{y}$.  The  associated  two-curvature 
also has only one non-zero component, $\bar{R}_{xy}^{~~12}
(\bar{\omega})=-\frac{1}{y^2}$, and the scalar two-curvature is 
a negative constant:
$\bar{R}(\bar{\omega})=-\frac{2}{\ell^2}$. For  the choice $M_a^i=\lambda 
e_a^i,~\bar{M}_a^i=\bar{\lambda}e_a^i$ and proceeding as in the previous two 
examples, we obtain the connection one-forms:
\begin{eqnarray*}
\omega^{13}&=&\frac{\ell}{y}\left[\lambda dx+\phi 
dy\right],~~~~~~~~~\omega^{23}=\frac{\ell}{y}\left[-\phi dx+\lambda dy\right]~,\\
\omega^{14}&=&\frac{\ell}{y}\left[\bar{\lambda} 
dx+\bar{\phi}dy\right],~~~~~~~~\omega^{24}=\frac{\ell}{y}\left[-\bar{\phi} 
dx+\bar{\lambda} dy\right],~\\
\omega^{12}&=&\left[N-\frac{1}{y}\right] dx+\bar{N} dy,~~~\omega^{34}=0~~.
\end{eqnarray*}

\section{Conclusion}

We have constructed a complete space of solutions of the first order theory of gravity for degenerate tetrads with two zero eigenvalues. This 
space is spanned by an infinite number of degenerate four-geometries.
Associated connection fields generically contain torsion and hence these 
configurations can be viewed as  geometric sources of torsion in pure 
gravity.  It should be emphasized that even though it is always possible to find (locally) a two-dimensional reduced subspace encoded by the diads $e_a^i$ for each configuration, these are four dimensional solutions described by the set of arbitrary fields $M^{kl}$, $~\bar{M}^{kl}$, $~\phi$, 
 $~\bar{\phi}$, $~N$ and $~\bar{N}$ which  depend on all the four spacetime 
 coordinates. The space of solutions is shown to contain a special 
class associated with the three independent homogeneous two-geometries of 
constant curvature, namely, $E^2,~S^2$ and $H^2$. This provides a  
geometric classification of this family of infinitely many degenerate spacetime solutions of first order gravity in four dimensions. 
 
The present study complements the earlier  analysis for tetrads with   one null 
eigenvalue \cite{kaul}.  Altogether, these configurations
described by tetrads with  one, two and three zero eigenvalues  constitute 
all possible degenerate solutions of the equations of 
motion in first order gravity. 

Let us note that one could also include the cosmological constant in this analysis. However, for the case of tetrads with two or more zero eigenvalues as discussed here, its inclusion does not affect any of the connection or tetrad equations of motion. This is in stark contrast to the case with one null eigenvalue, where the addition of the cosmological constant does affect the tetrad equations of motion, and hence also the solutions. 

The degenerate spacetime solutions constructed here, as well as those found earlier 
\cite{kaul}, are potential mediators of topology (signature) change. 
As  saddle points in the  quantum path integral, these  could also encode 
nontrivial quantum effects. In particular, these might underlie 
hitherto unnoticed but interesting instanton phenomena. 
These questions need to be explored for a deeper understanding of 
quantum gravity. 

\acknowledgments
S.S. thanks Sayan Kar for useful discussions.
R.K.K.   acknowledges the support of Department
of Science and Technology, Government of India, through a J.C.
Bose National Fellowship.

\end{document}